\documentclass[a4paper,11pt]{article}

\usepackage{pos}
\usepackage{lipsum} 
\usepackage{lineno}

\title{Probing neutrino emission at GeV energies from astrophysical transient events with the IceCube Neutrino Observatory}

\ShortTitle{GeV neutrinos in IceCube}

\author{The IceCube Collaboration \\{\normalsize \normalfont(a complete list of authors can be found at the end of the proceedings)}\\}

\emailAdd{gwenhael.dewasseige@uclouvain.be}
\emailAdd{karlijn.kruiswijk@uclouvain.be}

\abstract{

Astrophysical transient events like Gamma Ray Bursts (GRBs) have always been promising candidates for multi-messenger astronomy, with electromagnetic and gravitational wave signals having already been observed in GRBs such as GRB 170817A. The neutrino signatures of these bursts have been long-awaited as well, with many models predicting different spectra. Most of these searches have been in the hundreds of GeV to PeV range. However, as different models indicate a possible lower energy neutrino signal, we intend to expand this search to the lowest limits of IceCube (0.5-5 GeV) as well. With the plan to look at more transient events, we present the result of the first IceCube search for < 5 GeV astrophysical neutrinos emitted from a GRB, for GRB 221009A; the brightest GRB ever observed. Furthermore, we present plans to improve the observations of < 5 GeV neutrinos in IceCube, with which we plan to probe more transient events in the future. These improvements include the addition of direction reconstruction at these energies, and optimization of the noise rejection. With these improvements, GRB 221009A is just the start of the low-energy neutrino search from transient events with IceCube.

\vspace{4mm}
{\bfseries Corresponding authors:}
Gwenhaël de Wasseige$^*$,Karlijn Kruiswijk\\
{ \itshape Centre for Cosmology, Particle Physics and Phenomenology - CP3, Universit{\'e} catholique de Louvain, Louvain-la-Neuve, Belgium}\\[4mm]
$^*$ Presenter

\ConferenceLogo{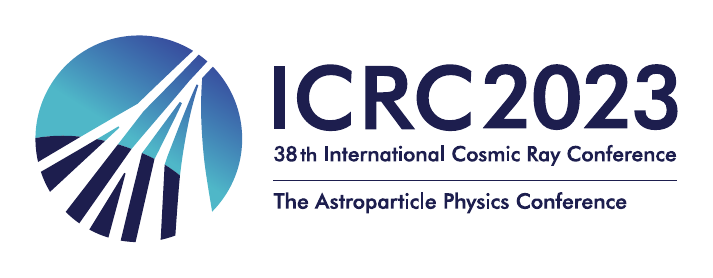}

\FullConference{The 38th International Cosmic Ray Conference (ICRC2023)\\ 26 July -- 3 August, 2023\\ Nagoya, Japan}
}

\begin{document}

\maketitle

\section{Introduction: GeV neutrinos in multi-messenger astronomy}\label{secIntro}
In multi-messenger astronomy, the neutrino is particularly interesting. With its minuscule cross-section and neutral electric charge, it is able to pass through magnetic fields, gas clouds and other barriers that would stop other messengers. Therefore the observed direction of neutrinos will match the direction of the neutrino source. It is even able to cross the outer layers of the source itself, allowing a new view inside astrophysical objects. Moreover, neutrinos indicate the presence of hadronuclear reactions that could not be seen with conventional astronomy. These neutrino properties aid to further research into the physics and the elements present in distant astrophysical processes. 

Neutrinos can be created across a range of energies, by different processes and sources. The first detected astrophysical source of neutrinos was the sun. These solar neutrinos originate from the sun's inner core through nuclear reactions, and have below MeV energies~\cite{solarneutrinos}. In the TeV--PeV range, IceCube~\cite{IceCube:2016zyt} has been able to find evidence of astrophysical sources, namely TXS~0506+056~\cite{TXS} and  NGC~1068~\cite{NGC1068}. These neutrinos are expected to come from proton-photon interactions. However, these named sources are just the tip of the iceberg, as there are many possible astrophysical sources that could be able to emit neutrinos. 
IceCube is also capable of observing neutrinos at GeV energies. These neutrinos come from proton-proton and proton-neutron collisions, and have the possibility to be created in sources where these protons and neutrons can be accelerated. Gamma Ray Bursts (GRBs) are a popular candidate for these processes, where neutrinos can be created below the photosphere~\cite{Murase:2013hh}.

Nowadays, there are several neutrino detectors looking for astrophysical neutrinos, including IceCube, Super-Kamiokande and KM3NeT. Super-Kamiokande is currently the only fully operating detector that is specifically optimized for MeV to GeV
neutrinos, and has been able to provide upper limits on astrophysical neutrinos in the GeV range~\cite{Abe_2018}.
However, this detector has a limited size of 50~kt of water, making it incredibly difficult to further improve its sensitivity. Instead, a bigger detector is preferred to probe the fainter astrophysical sources. The largest operational neutrino detector at this moment is IceCube, whose km$^3$ detector volume allows it to detect faint astrophysical neutrino sources. This detector, made up of strings of Digital Optical Modules (DOMs), each with a Photo Multiplier Tube (PMT), is optimized for the TeV--PeV range of neutrino energies, but is also capable of observing GeV neutrinos. KM3NeT will be another contender for detector size, once finished and includes a part optimized for GeV neutrinos. However, since it is not completed, no astrophysical sources have been observed.

We can observe GeV neutrinos in IceCube using DeepCore, a more densely instrumented volume of the detector containing PMTs more sensitive than the standard IceCube ones.
In this energy range most observed neutrinos come from the atmosphere, which on their own are interesting, e.g to study neutrino oscillations, but still hinder the astrophysical neutrino search. However, it is still possible to probe the GeV astrophysical neutrinos. By comparing the number of observed neutrinos during a transient astrophysical event to that of a transient-less period, a new part of the neutrino spectrum can be used to discover new properties of astrophysical sources. 

\section{IceCube GeV neutrino research}\label{secneutrino} 
With DeepCore, it is possible to look for GeV neutrinos, where with a specialized selection called ELOWEN an energy range of 0.5 to 5 GeV can be reached~\cite{Abbasi_2021GeVSolar}. This selection removes higher energy events, as well as detector noise, the largest source background for this selection, through several different filtering steps. The end result allows for astrophysical GeV neutrino research. 

The ELOWEN selection lowers the initial 1400 Hz of data down to 0.02 Hz, which is explained more thoroughly in~\cite{Abbasi_2021GeVSolar}.
The initial data for this filter selection is dominated by atmospheric muons and "noise events", caused by the detector itself and including uncorrelated thermal noise, uncorrelated radioactive noise, and correlated scintillation noise~\cite{Abbasi_2010,ABBASI2009294}. 
To remove this noise, several filtering steps are taken. 
The first step is to look solely at events that appear inside DeepCore, and do not trigger any other filters specialized for higher energy events, which reduces the noise rate to 15Hz. The next step to remove this noise is a constraint on the number of triggered optical modules, reducing the data to 6 Hz. This is to further remove higher energy events and events that are not fully contained in DeepCore, as well as detector noise. 
Another step considers the amount of correlated hits, using the NoiseEngine filtering step~\cite{MLarson}. This reduces the rate of detector noise events to 0.2 Hz. 
Lastly, data quality selections on estimated depths of the interaction point, total charge of the event, and the distance and delay between the first two event hits help reduce the data rate down to 0.02 Hz. This final sample contains 40\% of the initial sample of neutrinos following a standard $ E^{-2}$ spectrum from 0.5 to 5 GeV triggering the detector, simulated with GENIE 2.8.6~\cite{ANDREOPOULOS201087}. The final sample is still dominated by the remaining noise events and is larger than the expectation for atmospheric neutrinos, which are estimated to occur at the mHz level.

With this filter it is possible to look for transient events, even with a large background and lack of directional reconstruction. Because of the nature of transients, only the number of neutrino events during the transient event has to be known, which can then be compared to the background in an off-time region to provide its significance. 
With this method, the ELOWEN selection has been targeting neutrinos from solar flares~\cite{Abbasi_2021GeVSolar}, as well as from binary mergers~\cite{abbasi_2021GeVmergers}~\cite{posterO4}, and even the brightest Gamma Ray Burst ever observed. 

\subsection{GRB221009A}\label{ssecGRB}
The search for GeV neutrinos from the Brightest Of All Time GRB 221009A~\cite{Boat}, was done together with different analyses, ranging from MeV--PeV~\cite{GRB22paper}~\cite{TalkGRB22}. In this search, ELOWEN looked for neutrinos in two time windows, one 1000 s time window centered on the initial detection time T$_0$, and a 2200 s time-window which started 200 s before the initial detection time and was shared between the different energy ranges.

The background sample for this search was made up of time windows during which no transient events were observed, including novae, GRB, gravitational waves from merger events, and solar flares. For each different duration a background sample of 2600 of these time windows was created as a comparison.

To ensure the quality of the data taking during the GRB, several checks were performed. Prior to the unblinding of the data, in addition to the automated data quality checks performed continuously in IceCube~\cite{Aartsen_2017}, an 8 hour time window just before the chosen time window is used to evaluate the compatibility of the detector rate with the expected 0.02\,Hz background data rate without any unexpected statistical fluctuations. Furthermore, the distribution of the neutrino candidate events during the 1000 s and 2200 s time windows are checked to exclude any potential neutrino candidate events to be caused by detector effects at a specific location.

\begin{figure}[hbtp]
    \centering
    \includegraphics[width=0.6\textwidth]{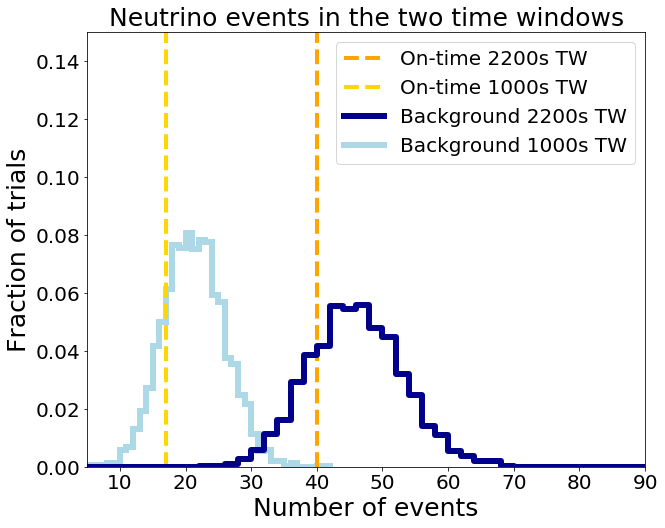}
    \caption{The result of the 1000s (yellow) and 2200s (orange) time window search during GRB 221009A with ELOWEN, compared to their expected background in light and dark blue respectively. The observations during the GRB searches comply with their background expectation.}
    \label{figGRB221009}
\end{figure}

In the time window of GRB221009A, all checks performed passed. The result of these searches is visible in Figure \ref{figGRB221009}. In both cases there is no significant deviation from the expected background, with p-values of 0.79 for the 1000s window and 0.81 for the 2200s window. Using the effective area of ELOWEN, we can then derive a 90\% C.L. upper limit per time window on the time-integrated all-flavor neutrino fluence and flux (for a reference energy of 1~GeV), assuming a power-law emission with a spectral index of 2 between the energies of 0.5 and 5~GeV. The upper limit derived is $5.3\times10^3$~GeV~cm$^{-2}$ for the 1000~s time window and $7.9\times10^3$~GeV~cm$^{-2}$  for the 2200~s time window.
The differential limits and model exclusions calculated from this search, together with other analyses can be found in~\cite{TalkGRB22}.

\subsection{Binary mergers}\label{ssecGW}
Several searches for neutrinos from binary mergers have been done, with complete catalogue searches up to O3~\cite{abbasi_2021GeVmergers} and the current search for neutrino messengers concurring with gravitational waves from O4 still going on~\cite{posterO4}. 
For these events, the standard time window of 1000s is used, with the same background sample as for GRB 221009A.
Furthermore, in the case of binary neutron star mergers, an alternate three-second time window is also analyzed to search for neutrinos from possible precursors.
In both cases, upper limits are set on the neutrino flux, and further information can be found in~\cite{posterO4}.

\section{Improvements to ELOWEN}\label{secImprovements}
The biggest background in the ELOWEN selection is still detector noise: uncorrelated thermal noise, uncorrelated radioactive noise, and correlated scintillation noise~\cite{Abbasi_2010,ABBASI2009294}. Much of this noise is already reduced by the many different filtering stages, but there are improvements to be made, especially by applying machine learning. 

With the help of different machine learning algorithms, it is possible to further filter down the noise level, improving the ELOWEN sensitivity. There is also the possibility to reconstruct the direction of some neutrino events, something that previously was not thought possible at single GeV energies.

\subsection{Noise filtering}\label{ssecNoise}
The current filtering steps with ELOWEN work well to reduce the noise while preserving the low-energy neutrinos, allowing us to constrain the low-energy neutrino flux. However, there is still room for improvements in this noise filtering process.
By reevaluating all the different variables used in each filtering steps, especially when combined together, a gain in the signal to noise ratio can be obtained. Furthermore, one can allow some filters to be specifically trained to improve its filtering capabilities with the help of machine learning, by training on known neutrino and noise simulations.

To reevaluate the different variables used in the filtering steps, it is important to look at them as a whole. To do this, we use an unsupervised machine learning algorithm called t-SNE to examine whether the full combination of all variables used in the filtering steps can show further differences between noise and neutrinos. 
The t-SNE algorithm is able to reduce the dimension of the parameter space, taking into account the clustering of the events in that parameter space. The algorithm creates a Student's t-distribution for the distance to each event based on the density of the events surrounding it, and gives a probability based on this distribution for each pair of events. Here similar events have a high probability. The algorithm then creates a similar distribution in the two-dimensional space, and fits the positions of each event in this space according to the distributions. Though the resulting positions in the 2D space are not exactly the same every time, similar events will always cluster together.

\begin{figure}[hbtp]
\begin{minipage}[t]{.48\linewidth}
    \centering
    \includegraphics[width=\textwidth]{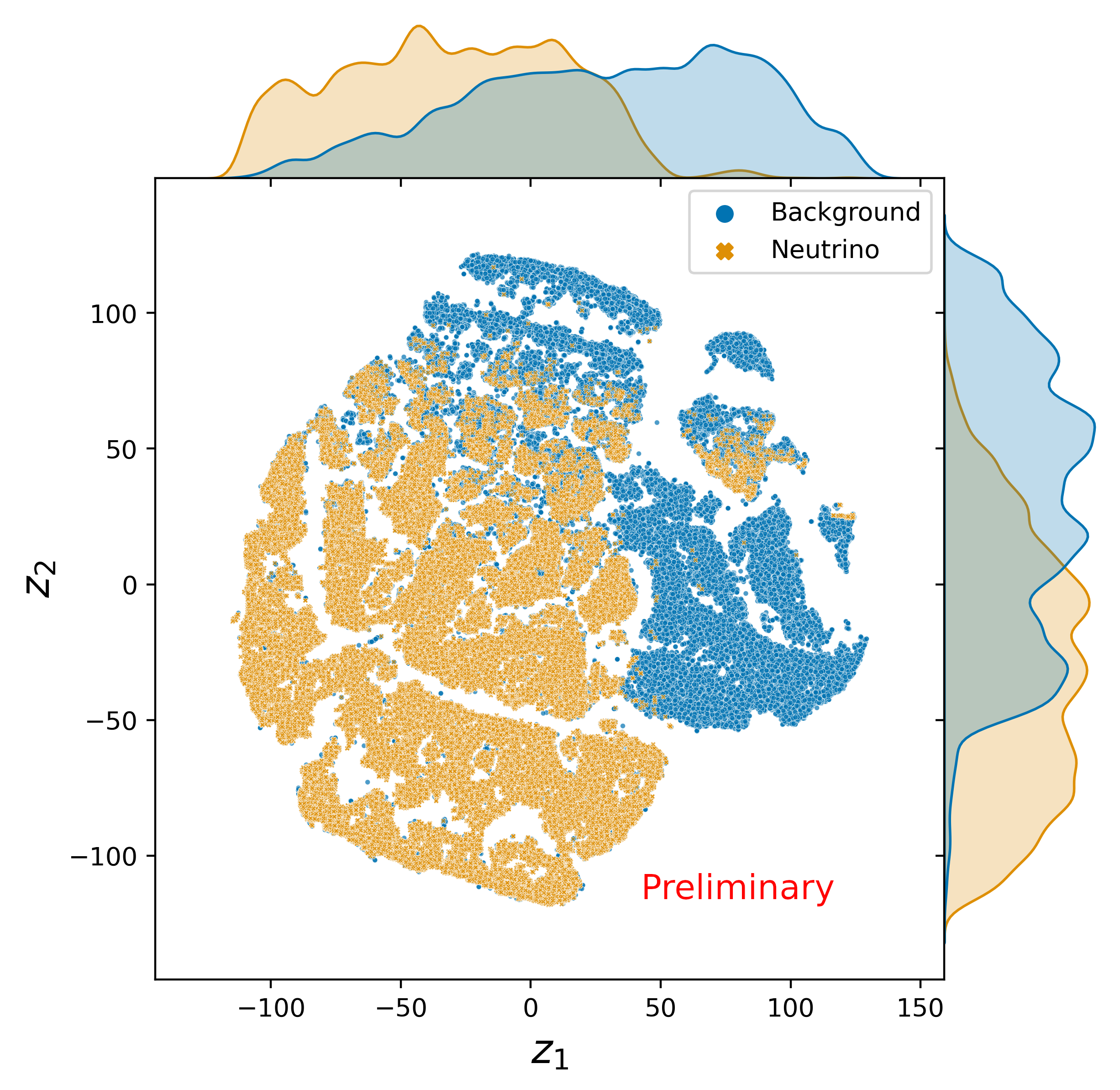}
    \caption{A t-SNE plot of the data from neutrino simulations and background, with $Z_1$ and $Z_2$ arbitrary parameters specifying distance between points. The histograms of the distributions in $Z_1$ and $Z_2$ are shown on the top and right sides. Visible are some clusters (upper right) are completely dominated by background, while others seem to consist mostly of simulated neutrinos (lower left). }
    \label{ftsne}
\end{minipage}\centering \quad
\begin{minipage}[t]{.48\linewidth}
    \centering
    \includegraphics[width=\textwidth]{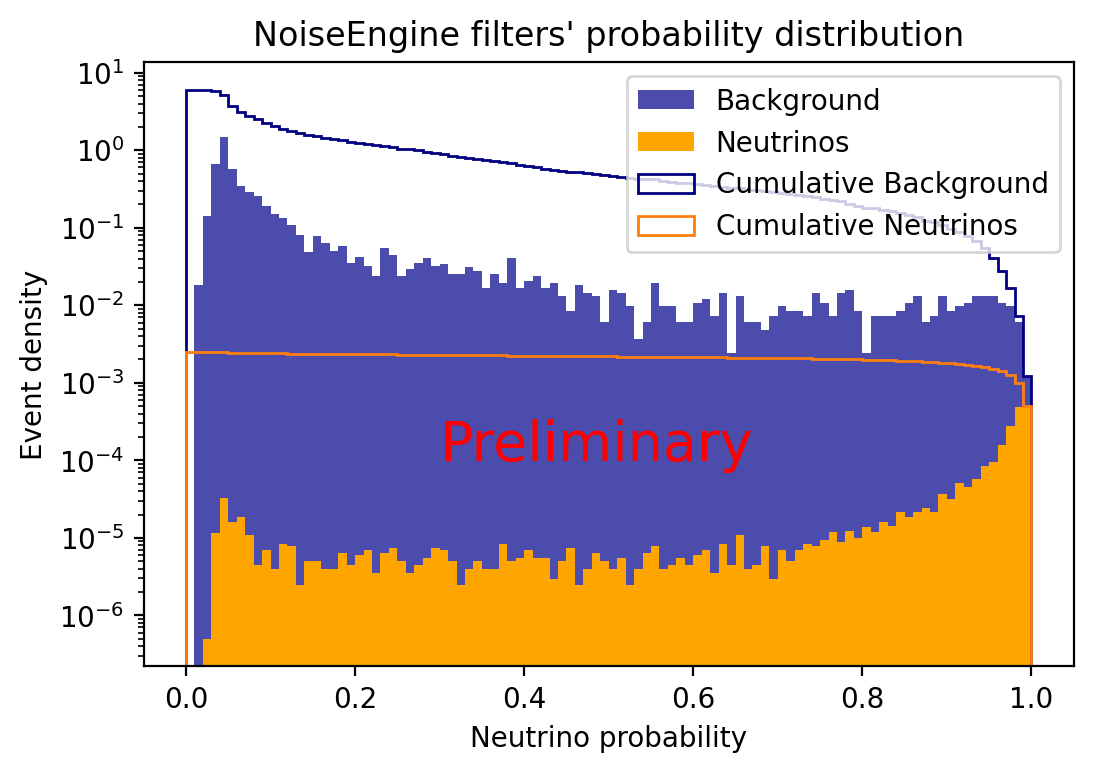}
    \caption{A plot of the probability distribution of neutrino and noise events and the BDT's neutrino probability score. Both the neutrino and the background densities are scaled to show the ratio of their rates at this filtering step. The rate of neutrinos and noise passing is similar to each other for very high neutrino probabilities (>0.95).}
    \label{fnprob}
\end{minipage}
\end{figure}

The resulting t-SNE plot of the filtering variables can be seen in Figure \ref{ftsne}. Here, the background is made up of real data during times where no transient events were detected, and the neutrinos are simulations in the energy range of 0.5 -- 5 GeV. 
In this case, the input data consists of low level variables used in the filtering steps, see~\cite{Abbasi_2021GeVSolar}. 
From this t-SNE plot, it is clear that while it shows a lot of overlap between background and neutrinos around $Z_1 =0,Z_2 =0$, there are several clusters which are dominated by neutrinos. This shows that the combination of different filtering steps does allow for good separation between noise and neutrinos.

Another way to improve some filtering steps is to work on the individual filtering steps' ability to distinguish neutrinos from background. This was done with the NoiseEngine filtering step, which looks for correlated hits~\cite{MLarson}. Specifically, NoiseEngine searches how many pairs of hits fit in a specific apparent velocity and time window, and can select events where the required amount of pairs is met. Currently, a combination of 8 different NoiseEngine settings for different time and velocity windows is used, but the filtering can be improved by using more of these different settings, and combining them with the use of machine learning.

With a boosted decision tree (BDT), a clear distinction can be made between noise and neutrinos. However, because of the relatively large amount of noise present compared to the low number of neutrinos expected, the filter needs to be able to strongly reduce the noise while keeping enough neutrinos. The BDT therefore has to be optimized for the signal to noise ratio instead of the accuracy. The resulting probability distribution of the BDT, scaled to the expected ratio of the background and neutrino rates, can be found in Figure \ref{fnprob}. The boosted decision tree has a high accuracy, and by ensuring a high threshold (>0.95) the noise can be brought down to a level similar to the neutrinos.

With the combination of creating more strict thresholds of filters and joining several filters together, a great improvement can be made to the noise filtering. Joining filters together permits more precise filtering, and the ability to observe neutrinos that would have previously been hidden to us. The use of machine learning allows us to make these filters more precise and more strict, reducing the amount of background noise and improving the detection capabilities and sensitivity.

\subsection{Direction reconstruction}\label{ssecDirection}
Another area where a big improvement can be made is in the direction reconstruction, as there is presently no direction reconstruction available at GeV energies. This is mostly because at these energies the neutrino event is small in comparison to the distances between the individual PMTs of IceCube, and even those of DeepCore: for single GeV neutrinos muons have a track length of up to 24 m, while the horizontal spacing inside DeepCore is 72 m. However, the zenith direction of the neutrinos can be reconstructed, as the DOM spacing on a DeepCore string is 7 m, which is more similar to the expected track length in the event.

For reconstructing the zenith, the hits on a single string can be used. Taking the string with the most hits, the zenith direction of the neutrino can be estimated by comparing the hits on the surrounding 7 DOMs, including the DOM with the first Hard Local Coincidence (HLC) hit. The  downward facing DOMs allow upward travelling neutrinos to be detected with more hits than downwards travelling neutrinos. Furthermore, there is a difference in timing between the first hit and the hits on the DOMs below/above for upwards and downwards neutrinos that can be capitalized on. 

\begin{figure}[hbtp]
    \centering
    \makebox[\textwidth][c]{\includegraphics[width=1.3\textwidth]{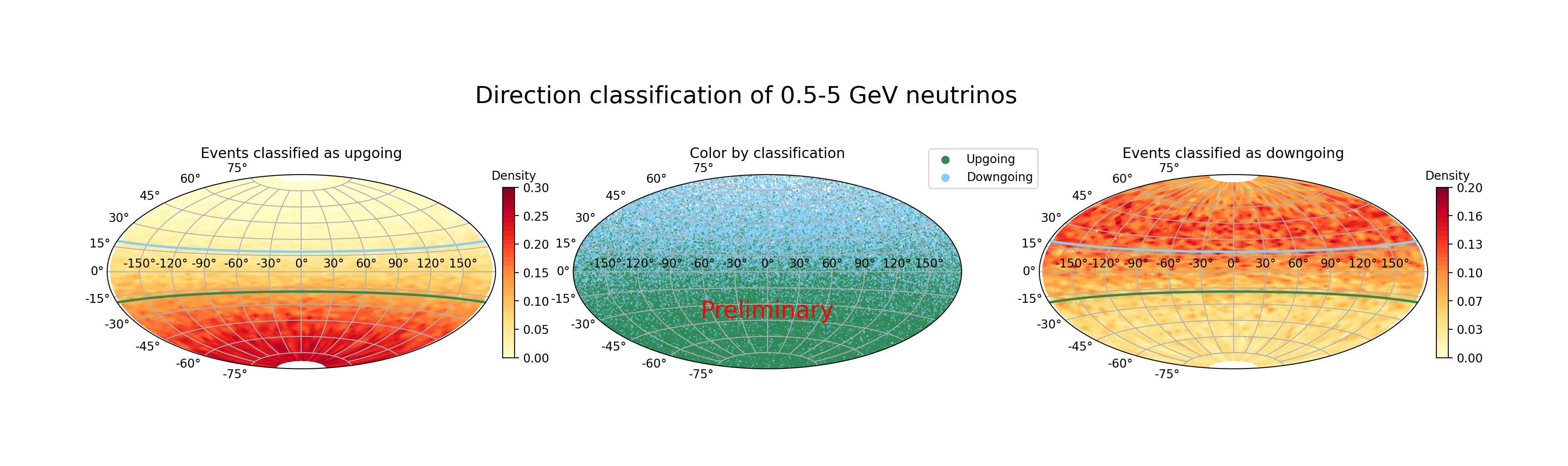}}
    \caption{The direction reconstruction with the use of two boosted decision trees. The middle panel shows the category of the simulated events, with the left and right panels showing the densities of the up- and downgoing neutrinos, respectively. The blue and green lines belong to the angles used for training the up- and downgoing neutrinos. }
    \label{fdirections}
\end{figure}

Using a combination of two boosted decision trees, trained on recognising upwards and downwards travelling neutrinos respectively, a balanced accuracy (corrected for the unbalance in catagories) of 77\% can be reached, of which the resulting classification of each event can be seen in Figure \ref{fdirections}. Here 40\% of neutrino events are not recognized as either up or down, either because they are horizontal, or consist of too few hits.
With the use of more advanced reconstruction algorithms, an even better reconstruction of the zenith direction can be achieved. This is currently being worked on, to see if more neutrino events can be reconstructed with a better accuracy.

The use of even a rough direction reconstruction is very useful, as it would improve the signal to noise ratio. Furthermore, it allows us to better search for neutrinos from astrophysical sources by using the direction reconstruction. 

\section{Conclusions and prospects}\label{secCon}
To conclude, IceCube is capable of observing 0.5 to 5~GeV neutrinos, and is actively doing transient searches for neutrinos in this energy range. Furthermore, there is active research to improve IceCube's capabilities in this energy range. First of all, there are efforts continuing to decrease noise and improve the sensitivity, which appear promising. Secondly, the possibility to add direction reconstruction in the zenith direction has been developed, which is the first in this energy range for IceCube. With these improvements, IceCube will be more sensitive to neutrinos in the GeV energy range, with which there are plans to continue searches for neutrinos from transient events to better our understanding of the universe and its contents.

\bibliographystyle{ICRC}
\bibliography{references}

%

\clearpage

\section*{Full Author List: IceCube Collaboration}

\scriptsize
\noindent
R. Abbasi$^{17}$,
M. Ackermann$^{63}$,
J. Adams$^{18}$,
S. K. Agarwalla$^{40,\: 64}$,
J. A. Aguilar$^{12}$,
M. Ahlers$^{22}$,
J.M. Alameddine$^{23}$,
N. M. Amin$^{44}$,
K. Andeen$^{42}$,
G. Anton$^{26}$
% ...
\\
\\
% ...
$^{12}$ Universit{\'e} Libre de Bruxelles, Science Faculty CP230, B-1050 Brussels, Belgium \\
% ...
$^{17}$ Department of Physics, Loyola University Chicago, Chicago, IL 60660, USA \\
$^{18}$ Dept. of Physics and Astronomy, University of Canterbury, Private Bag 4800, Christchurch, New Zealand \\
% ...
$^{22}$ Niels Bohr Institute, University of Copenhagen, DK-2100 Copenhagen, Denmark \\
$^{23}$ Dept. of Physics, TU Dortmund University, D-44221 Dortmund, Germany \\
% ...
$^{26}$ Erlangen Centre for Astroparticle Physics, Friedrich-Alexander-Universit{\"a}t Erlangen-N{\"u}rnberg, D-91058 Erlangen, Germany \\
% ...
$^{40}$ Dept. of Physics and Wisconsin IceCube Particle Astrophysics Center, University of Wisconsin{\textendash}Madison, Madison, WI 53706, USA \\
% ...
$^{42}$ Department of Physics, Marquette University, Milwaukee, WI, 53201, USA \\
% ...
$^{44}$ Bartol Research Institute and Dept. of Physics and Astronomy, University of Delaware, Newark, DE 19716, USA \\
% ...
$^{63}$ Deutsches Elektronen-Synchrotron DESY, Platanenallee 6, 15738 Zeuthen, Germany  \\
$^{64}$ Institute of Physics, Sachivalaya Marg, Sainik School Post, Bhubaneswar 751005, India \\
% ...

\subsection*{Acknowledgements}

\noindent
The authors gratefully acknowledge the support from the following agencies and institutions:
USA {\textendash} U.S. National Science Foundation-Office of Polar Programs,
U.S. National Science Foundation-Physics Division,
U.S. National Science Foundation-EPSCoR,
Wisconsin Alumni Research Foundation,
Center for High Throughput Computing (CHTC) at the University of Wisconsin{\textendash}Madison,
Open Science Grid (OSG),
Advanced Cyberinfrastructure Coordination Ecosystem: Services {\&} Support (ACCESS),
Frontera computing project at the Texas Advanced Computing Center,
U.S. Department of Energy-National Energy Research Scientific Computing Center,
Particle astrophysics research computing center at the University of Maryland,
Institute for Cyber-Enabled Research at Michigan State University,
and Astroparticle physics computational facility at Marquette University;
Belgium {\textendash} Funds for Scientific Research (FRS-FNRS and FWO),
FWO Odysseus and Big Science programmes,
and Belgian Federal Science Policy Office (Belspo);
Germany {\textendash} Bundesministerium f{\"u}r Bildung und Forschung (BMBF),
Deutsche Forschungsgemeinschaft (DFG),
Helmholtz Alliance for Astroparticle Physics (HAP),
Initiative and Networking Fund of the Helmholtz Association,
Deutsches Elektronen Synchrotron (DESY),
and High Performance Computing cluster of the RWTH Aachen;
Sweden {\textendash} Swedish Research Council,
Swedish Polar Research Secretariat,
Swedish National Infrastructure for Computing (SNIC),
and Knut and Alice Wallenberg Foundation;
European Union {\textendash} EGI Advanced Computing for research;
Australia {\textendash} Australian Research Council;
Canada {\textendash} Natural Sciences and Engineering Research Council of Canada,
Calcul Qu{\'e}bec, Compute Ontario, Canada Foundation for Innovation, WestGrid, and Compute Canada;
Denmark {\textendash} Villum Fonden, Carlsberg Foundation, and European Commission;
New Zealand {\textendash} Marsden Fund;
Japan {\textendash} Japan Society for Promotion of Science (JSPS)
and Institute for Global Prominent Research (IGPR) of Chiba University;
Korea {\textendash} National Research Foundation of Korea (NRF);
Switzerland {\textendash} Swiss National Science Foundation (SNSF);
United Kingdom {\textendash} Department of Physics, University of Oxford.

\end{document}